# Designing More Engaging Serious Games to Support Students' Mental Health: A Pilot Study Based on A CBT-Informed Design Framework

**Ting-Chen Hsu** [1,*], **Zheyuan Zhang** [1], **Ziyi Chen** [1], **Yuwen Liu** [1], **Yanjia Liu** [1]

[1] School of Animation and Digital Arts, Communication University of China, Beijing, China
[*] Corresponding Author: Ting-Chen Hsu (e-mail: tingchenhsu.ac@gmail.com)

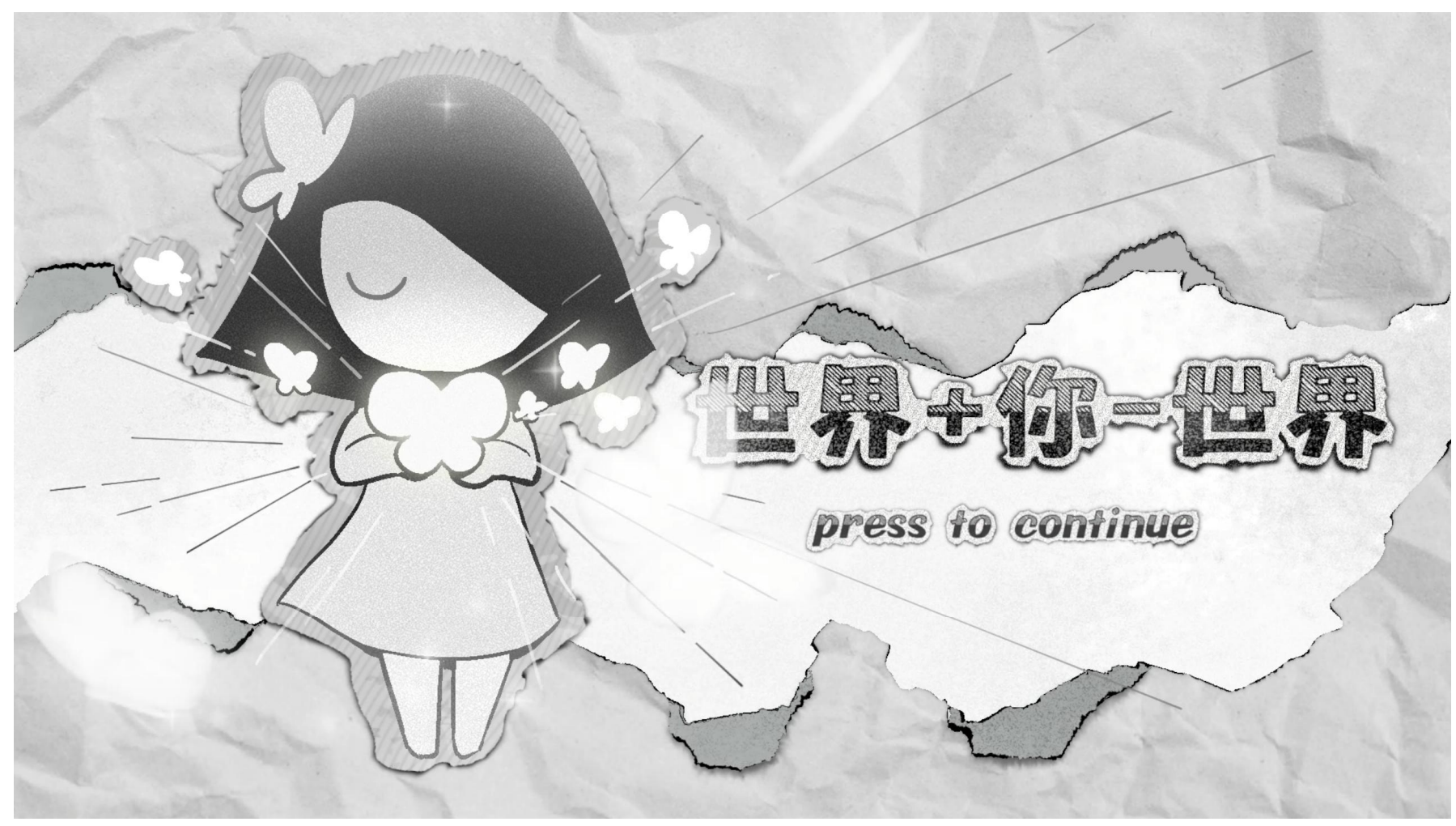

**Figure 1.** The cover of the serious game "World + You - World" designed based on the CBT-informed design framework "TPR-MMF".

**Abstract:** Addressing the issues of dullness, low compliance, and lack of appeal in current digital mental health education and serious games for students and adolescents, this study proposes a novel, experience-centered framework for serious game design—the Therapeutic Procedural Rhetoric and Mechanism Mapping Framework (TPR-MMF). Based on this framework, a side-scrolling serious game prototype, "World + You - World," was developed. This study compared the effectiveness of TPR-MMF-based games with traditional explicit educational serious games through a small-sample randomized controlled trial (N=28). The results of the Intrinsic Motivation Inventory (IMI) showed that the experimental group (playing "World + You - World") significantly outperformed the control group in four aspects. Furthermore, qualitative survey results indicated that players could perceive the psychological metaphors within the game mechanics and spontaneously resonated with real-life experiences. This study provides a highly engaging new development paradigm for gamified mental health education for students and adolescents.

## 1. Introduction

The incidence of mental health issues among students and adolescents is significantly increasing globally. A systematic review shows that approximately 25-31% of the student population is affected by mental health issues (Silva et al., 2020). Meanwhile, the global prevalence of depression or depressive symptoms among student populations is significantly increasing (Dongjun et al., 2025; Lu et al., 2024). Due to multiple factors such as academic pressure and peer relationships during adolescence, students become a large group that is easily troubled by psychological problems (Carvajal-Velez et al., 2023; Liu & Kuai, 2025). If these troubles are not prevented and intervened in a timely manner, they will affect learning, interpersonal relationships, and overall health development in adulthood (Schlack et al., 2021; Uhlhaas et al., 2023). In addition, due to insufficient psychological service resources and low willingness of students to learn about mental health knowledge, there is a serious lack of coverage of preventive measures for mental health issues (Benton et al., 2021, Michaud et al., 2020; Wahdi et al., 2025). Therefore, there is an urgent need to research more innovative solutions to the education of students' mental health.

Among existing evidence-based psychological interventions, cognitive behavioral therapy (CBT) is widely used for the prevention and treatment of mental health issues among students due to its clear theoretical structure and operational technical system (Christ et al., 2020; Zhang et al., 2025). In depression intervention, a meta-regression analysis covering 31 trials and 4335 participants indicated that CBT has a moderate level of effectiveness in preventing and improving student depression (Oud et al., 2019). In anxiety intervention, a meta-analysis showed that CBT is an effective means of preventing and improving anxiety disorders in children and adolescents (Pegg et al., 2022). However, the effectiveness and limitations of CBT coexist. On the one hand, CBT emphasizes skills practice and homework, but insufficient adherence can weaken skill mastery and consolidation (Jungbluth & Shirk, 2013; Kazantzis, 2021). On the other hand, whether face-to-face or online CBT, the initiation of therapy, sustained participation, and dropout rates are key bottlenecks affecting effectiveness evaluation and widespread application (Linardon et al., 2025). Meanwhile, for students, the attentional characteristics of their developmental stage and the perceived "boringness" of traditional psychoeducational materials may further increase mid-course dropout rates (Boucher & Raiker, 2024). Overall, although CBT at its current stage provides an important pathway for student mental health education, it still faces challenges in terms of sustainability and long-term effectiveness.

With the development of digital technology, the application of video games or gamification elements in mental health education has become a major research hotspot. Serious games based on cognitive behavioral therapy (CBT) hold promise for expanding the reach of mental health problem prevention and increasing students' participation (Fleming et al., 2017). For example, the gamified CBT project SPARX for depression showed comparable efficacy to conventional psychotherapy in randomized controlled trials (Fleming et al., 2021; Merry et al., 2012). MindLight, a serious game for anxiety that combines CBT elements with biofeedback mechanisms, has also been experimentally evaluated for its positive effects on preventing and improving anxiety symptoms (Schoneveld et al., 2016, 2020). The serious game "Silver" has demonstrated significant potential in enhancing cognitive regulation (De Jaegere et al., 2024). "Reverie", a dialogue based stress relief serious game, has also been shown to have preliminary effects in regulating mental health in pilot studies (Hsu, 2026). Some meta-analyses also show that although gamified mental health support has some shortcomings in terms of effectiveness, it has a positive overall effect (David et al., 2020; Townsend et al., 2022).

However, current research on serious games related to student mental health exhibits inconsistent results in terms of effectiveness, feasibility, and acceptability, with significant differences in research quality and assessment indicators (Gómez-León, 2025). Furthermore, many studies suffer from significant deficiencies in reporting on "design mechanisms," often lacking a clear theoretical framework to explain which specific game mechanisms modulate which CBT process variables and ultimately promote mental health literacy (Hammady & Arnab, 2022). A recent review of serious games for adolescents also emphasized the need for transparency and rigor in assessing game design features and user engagement (Zeiler et al., 2025). From another perspective, some scholars have proposed that gamification in education should be aligned with students' gaming preferences to enhance engagement and learning performance (García-Cabrera et al., 2025). Currently, most mainstream serious games primarily employ low-stimulation gameplay types such as narrative-driven progression, mini-game assembly, dialogue choices, and practice questions (Zeiler et al., 2025), while the mainstream game types preferred by students are mainly action, shooting, and puzzle genres (Shah, 2024). These two types differ significantly in core loop intensity.

Therefore, this study constructs a CBT-informed serious game design framework, TPR-MMF, centered on participatory experience. Based on this, the popular "side-scrolling platformer" genre (Li, 2019), which has been widely popular since the NES era, was selected, and a serious game prototype, "World + You - World," was designed and implemented. Finally, this study investigates the following questions through a small-sample randomized controlled trial and qualitative questionnaire testing:

· RQ1: Compared to explicitly educational serious games on mental health, does the TPR-MMF-based serious game enhance players' intrinsic motivation and experience quality?

· RQ2: Can players perceive the metaphors in the TPR-MMF-based serious game on mental health and generate reflection or real-world associations?

This pilot study will provide a CBT-informed serious game design paradigm, offering a new and feasible path for digital mental health education for students.

## 2. Methods

## 2.1. Framework Development

The core challenge in designing serious games aimed at students' mental health education lies in transforming mental health education into a highly engaging interactive experience. Because students tend to resist explicit, didactic instruction, this study developed a novel "Therapeutic Procedural Rhetoric and Mechanics Mapping Framework" (TPR-MMF). This framework, based on procedural rhetoric, abandons explicit mental health education. The TPR-MMF architecture is shown in Figure 2.

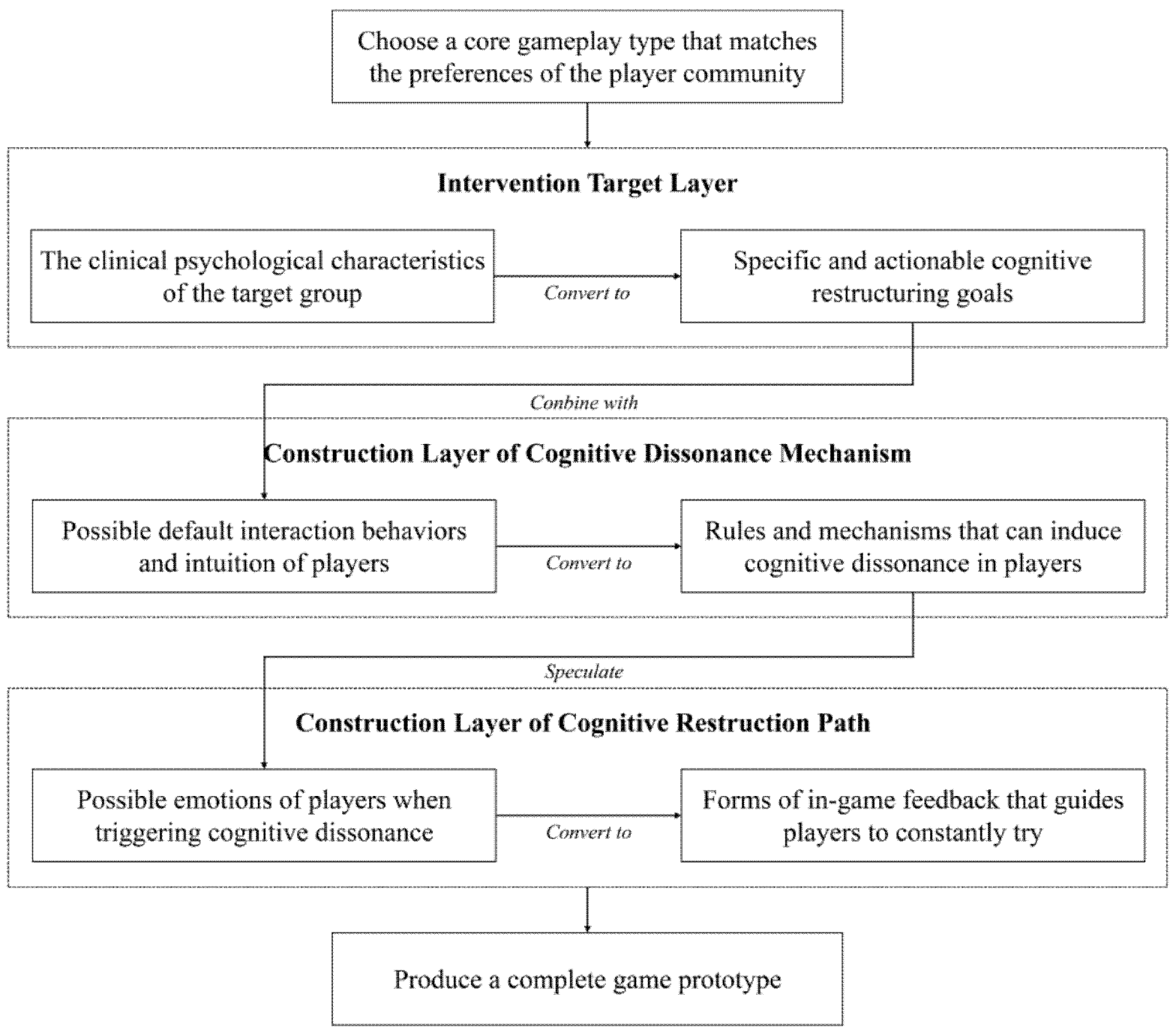

**Figure 2.** TPR-MMF schematic diagram

First, TPR-MMF emphasizes selecting gameplay types that align with target audience preferences for serious game design. On this basis, the first layer is the intervention target layer. This layer requires identifying the most vulnerable and destructive clinical psychological characteristics of the target group (e.g., perfectionist thinking) and translating them into specific and actionable cognitive restructuring goals (e.g., breaking the fixed notion that "only perfection equals success").

The second layer is the construction layer of cognitive dissonance mechanism. This layer combines the cognitive restructuring goals output from the first layer with players' likely default interaction behaviors and intuitions in a typical game environment, transforming them into specific game rules and mechanisms that deliberately induce "cognitive dissonance" in players. In this layer, the mechanism targets the intervention goal because the system deliberately blocks the player's default path to completion, triggering a strong "expectation violation." This experience forces players to break free from automated, distorted thinking patterns, providing psychological space for subsequent cognitive restructuring.

The third layer is the cognitive restructuring path construction layer. This layer requires inferring the player's emotions when cognitive dissonance is triggered based on the mechanisms that induce it, and establishing in-game feedback formats that guide players to continuously try different approaches. When a new strategy leads to successful completion of a level, this embodied "behavioral validation" can directly shake and reconstruct the original cognitive system.

## 2.2. Evaluation Study Design

### 2.2.1. Participants

To rigorously evaluate the effectiveness of serious mental health education games designed based on TPR-MMF and to specifically address RQ1 and RQ2, this pilot study employed a small-sample randomized controlled trial (RCT). A total of 28 first-year university students aged 18-19 (N=28) were recruited online and offline as participants. Inclusion criteria included four aspects: (1) familiarity with the basic operation of video games; (2) general stress and a PSS-10 test score of 14 or higher; (3) not diagnosed with major depressive disorder or currently taking psychiatric medications; and (4) not a psychology major or having conducted CBT theoretical research (making it difficult for them to experience the game as ordinary players). The 28 participants were then randomly assigned to the experimental group (N=14) and the control group (N=14). During the experiment, participants in the experimental group played the serious game “World + You - World” developed based on TPR-MMF. The control group played another serious game, NeuroQuest. NeuroQuest is a short, dialogue-driven role-playing game that emphasizes cognitive and decision-making exercises, helping students improve their emotional awareness and problem-solving abilities through a story-based approach. There were no significant differences between the two groups of participants in terms of age, sex, and baseline PSS-10 scores (Table 1). All participants signed informed consent forms before the start of the experiment, were informed of the research objectives, procedures, and data confidentiality, and had the right to withdraw from the experiment at any time.

**Table 1.** Basic Information of Participants

| | Experimental Group (N=14) | Control Group (N=14) | p |
|---|---|---|---|
| Gender (Male/Female) | 8/6 | 8/6 | 1.00 |
| Age (M±S.D.) | 18.14±0.36 | 18.21±0.43 | 0.63 |
| PSS-10 Score (M±S.D.) | 20.57±4.07 | 21±3.16 | 0.76 |

### 2.2.2. Measures

This study employed a mixed approach, combining data from scales and qualitative questionnaires to evaluate the effectiveness of games in mental health education and player experience. To answer RQ1, the study used the Intrinsic Motivation Inventory (IMI) (Appendix A.1), administered post-experiment, to evaluate differences between the two groups of players in subjective experience, engagement, and intrinsic motivation. The scoring system adopts the Likert scoring method (1-7 points). Participants in the experimental group also completed a specially designed online qualitative survey after playing the game. This survey included three questions for each game level:

· Q1: What was your first thought upon entering this level?
· Q2: What message do you think this level's design is trying to convey?
· Q3: Does this level remind you of any experiences in your real life?

These three questions respectively investigated participants’ “Exploring Motivation,” “Metaphorical Understanding,” and “Realistic Resonance.” Participants completed the three questions (15 in total) for five levels, drawing on their own experiences.

### 2.2.3. Procedures

The study was conducted in a quiet, well-lit, and undisturbed indoor room, with each participant’s session lasting approximately 30 minutes. During the pre-test phase, participants signed informed consent forms, filled out basic information, and completed the PSS-10 test. The intervention phase then commenced, with both groups of participants playing the experimental games "World + You-World" and NeuroQuest for 20 minutes, respectively. Immediately after the games, both groups completed the IMI. Finally, the experimental group completed an online qualitative survey.

### 2.2.4. Statistical Analysis

This study used Python (version: 3.12.5) to perform statistical analysis. All statistical tests were two-tailed tests, and the significance level was set at $\alpha=0.05$. For IMI measurement data, this study used independent samples t-test to compare the differences in motivation levels between the experimental group and the control group. Cohen’s d report effect size was used:

$$d=\frac{\bar{X}_E-\bar{X}_C}{s_p}, \tag{1}$$

where $\bar{X}_E$ is the sample mean of the experimental group and $\bar{X}_C$ is the sample mean of the control group. $s_p$ represents the pooled standard deviation:

$$s_p=\sqrt{\frac{(n_E-1)s_E^2+(n_C-1)s_C^2}{n_E+n_C-2}}, \tag{2}$$

where $n_E$ and $n_C$ are the sample sizes of the experimental group and the control group, respectively, and $s_E^2$ and $s_C^2$ are the variances of the experimental group and the control group, respectively.

Qualitative survey data employed thematic analysis. First, all texts were initially coded line by line. Then, the codes were aggregated at both the intra-level and cross-level levels to form themes, corresponding to three question dimensions: Q1 Exploring Motivation, Q2 Metaphorical Understanding, and Q3 Realistic Resonance. To improve reliability, three researchers independently coded the data before discussing and integrating the results. Finally, the results were summarized based on theme frequency and proportion, and a sunburst chart was used to display the theme distribution across each level.

# 3. Results

## 3.1. Game Degisn Results

### 3.1.1. Game Mechanisms Design

To enhance the attractiveness of the game, this study adopts the mature mechanism of "side-scrolling platformer" and designs a game prototype "World + You - World" based on the TPR-MMF, which combines entertainment and mental health promotion effects. This study selected five common types of cognitive distortions and completed the deconstruction and game mechanism transformation of the five core cognitive distortions through TPR-MMF (Table 2).

**Table 2.** Cognitive distortion and game mechanism transformation based on TPR-MMF

| Cognitive Distortions | Clinical Features | Game Metaphor design | Procedural Rhetoric |
|---|---|---|---|
| Perfectionism | Pursuing absolute perfection and equating imperfection with complete failure | Items that do not require complete collection | Moderate giving up does not mean failure. |
| Overgeneralization | Generalization a single negative event into an expectation of eternal failure | A platform that escapes as soon as players try to reach | A single setback does not predict the future. |
| Jumping to Conclusions | Negative future predictions without evidence | Some platforms that appear visually impossible to reach | The difficulties anticipated by the brain are often false. |
| Magnification | Exaggerating perceived threats and maintaining anxiety through avoidance behavior | Some obstacles that appear to be very large | Fear is often an amplified illusion. |
| Personalization | Attributing uncontrollable negative outcomes entirely to oneself | A predicament that cannot be overcome alone | The result is often the combined effect of multiple variables. |

### 3.1.2. Implementation of Game Levels

Finally, this study will use the Unity engine (2022.3.52f1c1, Windows 10) for game prototyping. Figure 3 shows the implementation results of each level in the game prototype.

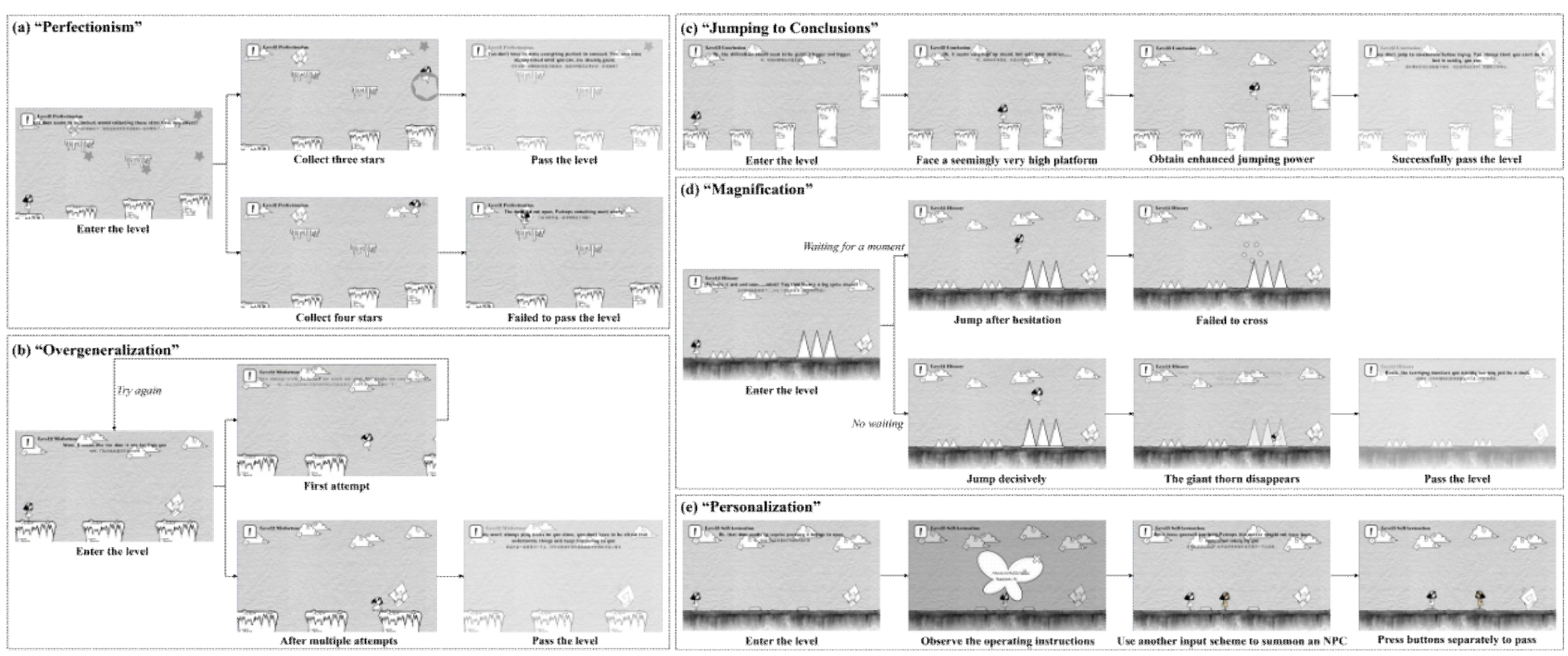

**Figure 3.** Implementation of Five Game Levels
**(a)** Flow of "Perfectionism" level. **(b)** Flow of "Overgeneralization" level. **(c)** Flow of "Jumping to Conclusions" level. **(d)** Flow of "Magnification" level. **(e)** Flow of "Personalization" level.

Perfectionism is an irrational thinking pattern that views things as "imperfect means failure". In the "Perfectionism" level (Figure 3.(a)), players can only pass the level smoothly by not collecting the most difficult star in the upper right corner. Stars represent "perfect completion" and "sense of achievement" in most level crossing games, but here we take the opposite approach and directly challenge players' inherent perception that perfection is equated with success.

Overgeneralization refers to summarizing a single negative event as an expectation of eternal failure. In the "Overgeneralization" level (Figure 3.(b)), when the player jumps to the platform on the right, the platform will suddenly move to the right. But as players try more times, the platform's movement distance will gradually decrease. In this level, the scene imposes random setbacks on players and encourages them to try and find solutions. After multiple attempts, players will realize that "failure is not eternal", gradually breaking the incomplete generalization of "eternal failure".

"Jumping to Conclusions" refers to making negative judgments without investigation or evidence, resulting in unnecessary negative emotions. In the "Jumping to Conclusions" level (Figure 3.(c)), the initially weakened jumping power and gradually increasing platform will make players subconsciously believe that they cannot achieve it. But as players keep jumping, they will find that their jumping power increases every time they jump to a new platform, and eventually they can reach the pass. The gradually increasing platform in the level prompts players to have a negative expectation of "this is impossible to complete". But every time the player tries, the level gives positive feedback that the jumping power is improved. Based on this, to help players reflect on their original negative expectations and gradually break the mindset of "Jumping to Conclusions".

"Magnification" refers to thinking about things in the worst direction, expecting the worst outcome, leading to extreme panic and helplessness. In the "Magnification" level (Figure 3.(d)), there is a giant spike on the right side of the scene. If the player hesitates repeatedly on the left side of the giant spike, they will not be able to cross it. On the contrary, if the player jump immediately after crossing the small spikes on the left, they will find that the giant spike is actually fake. The giant spike is a concrete manifestation of the idea of "Magnification". Player has a preconceived notion that "this is definitely impossible to jump over", repeatedly hesitating and ultimately leading to failure. But if the player take the initiative to jump decisively, they will realize that fear can be shattered.

"Personalization" refers to excessively attributing all responsibility to oneself, ignoring external factors or the responsibility of others. In the "Personalization" level (Figure 3.(e)), the player need to search for clues (press another set of directional keys) to summon an NPC, and control two characters to press buttons to open the door and pass the level. In many cases, when players are unable to complete levels independently, they subconsciously feel that it is their own problem. But in this level, by designing the collaboration of two characters, the player can realize that success and failure are not one-sided responsibilities, and thus correct the cognitive distortion of "always being solely responsible for everything".

## 3.2. Evaluation Results

### 3.2.1. Evaluation Results of IMI

Based on the IMI six-dimensional scores of 28 participants in the experimental group (E_Group, N=14) and the control group (C_Group, N=14), the differences in subjective experience between the two groups after the trial were compared using an independent samples t-test (Table 3).

**Table 3.** Analysis Results of IMI in various dimensions

| | Interest / Enjoyment | Perceived Competence | Effort / Importance | Pressure / Tension | Perceived Choice | Value / Usefulness |
|---|---|---|---|---|---|---|
| E_Group (M±S.D.) | 5.61±0.44 | 5.48±0.55 | 4.07±1.13 | 1.89±0.45 | 5.20±0.54 | 5.93±0.37 |
| C_Group (M±S.D.) | 4.78±0.52 | 4.79±0.65 | 4.19±0.51 | 1.70±0.45 | 4.58±0.40 | 5.35±0.30 |
| p | <.001*** | .005** | .732 | .290 | .002** | <.001*** |
| Cohen's d | 1.73 | 1.15 | -0.13 | 0.41 | 1.31 | 1.72 |

The results showed that the experimental group (E_Group) was significantly higher than the control group (C_Group) in the interest/enjoyment dimension (E_Group M=5.61, SD=0.44; C_Group M=4.78, SD=0.52; $p<.001$, Cohen's d=1.73) and also significantly higher in the perceived competence dimension (E_Group M=5.48, SD=0.55; C_Group M=4.79, SD=0.65; $p=.005$, d=1.15). Meanwhile, the experimental group also showed significant advantages in perceived choice (E_Group M=5.20, SD=0.54; C_Group M=4.58, SD=0.40; $p=.002$, d=1.31) and value/usefulness (E_Group M=5.93, SD=0.37; C_Group M=5.35, SD=0.30; $p<.001$, d=1.72), with relatively large effect sizes in all these areas. In comparison, there was no significant difference between the two groups in terms of effort/importance (E_Group M=4.07, SD=1.13; C_Group M=4.19, SD=0.51; $p=.732$, d=-0.13), nor was there a significant difference in the pressure/tension dimension (E_Group M=1.89, SD=0.45; C_Group M=1.70, SD=0.45; $p=.290$, d=0.41). Overall, the experimental group showed greater improvement in interest, sense of competence, sense of autonomy, and perceived value of the intervention content, supporting the advantages of the serious game "World+You-World" developed based on TPR-MMF in terms of attractiveness and participation experience.

### 3.2.2. Evaluation Results of Qualitative Questionnaire

Figure 4 illustrates the distribution of themes obtained from thematic analysis of 14 participants' answers to three types of questions (Q1 Exploring Motivation, Q2 Metaphorical Understanding, and Q3 Realistic Resonance) after the trial. A sunburst chart is used to present the main themes and their proportions for each level (L1-L5). Since the inner circle of Figure 4 assigns an equal weight of 20% to each level, and the outer circle represents the overall global proportion, for ease of explanation, the following text reports the level-specific proportion of each theme rather than the global proportion (i.e., the outer circle value multiplied by 5).

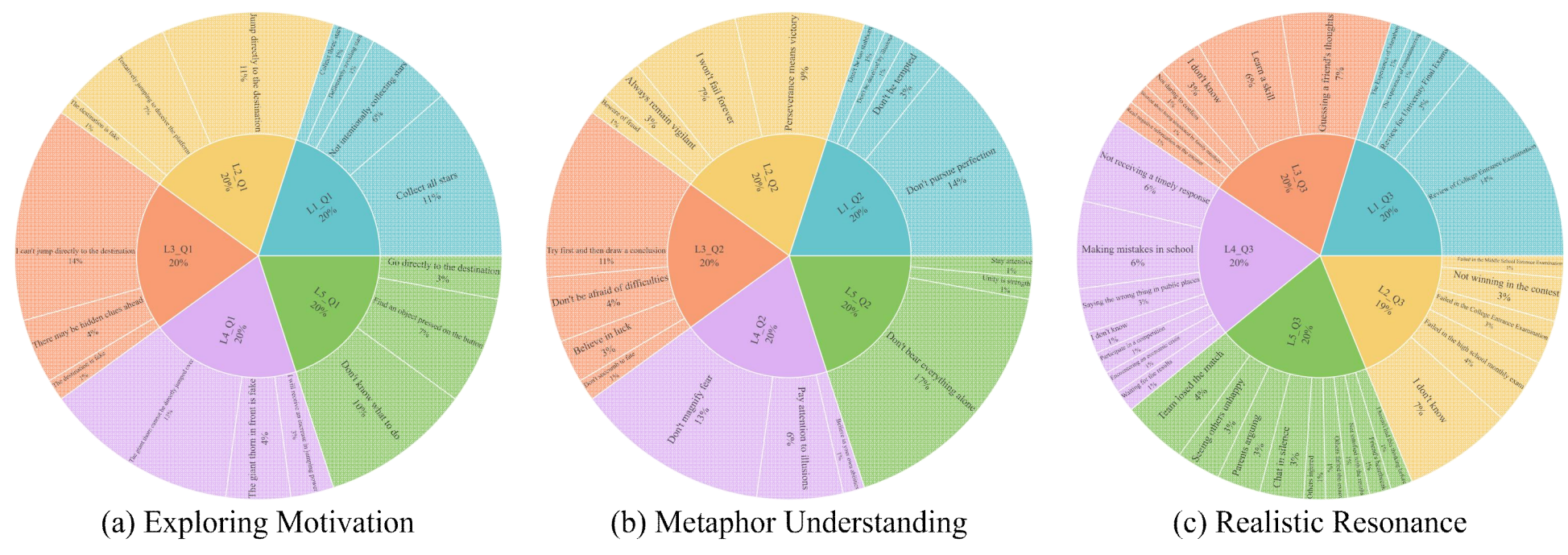

(a) Exploring Motivation (b) Metaphor Understanding (c) Realistic Resonance

**Figure 4.** Qualitative Questionnaire Evaluation Results of Three Dimensions
**(a)** Analysis results of "Exploring Motivation" dimension. **(b)** Analysis results of "Metaphor Understanding" dimension. **(c)** Analysis results of "Realistic Resonance" dimension.

As shown in Figure 4.(a), each level spontaneously activates a "default strategy" corresponding to its cognitive distortion. In the "Perfectionism" level, most participants first choose "collect all stars" (55%), reflecting an intuitive goal of equating "complete collection" with "success." In the "Overgeneralization" level, participants tend to choose "jump directly to the destination" (55%) or "tentatively jumping to deceive the platform" (35%), reflecting differences in their exploration of uncertain mechanisms. In the "Jumping to Conclusions" and "Magnification" levels, the immediate judgments of "can't jump directly to the destination" (70%) and "the giant thorn cannot be directly jumped over" (65%) are more prominent, indicating that difficulty and threat cues prioritize triggering negative expectations. The "Personalization" level shows "Don't know what to do" as the main reaction (50%), suggesting that there is a certain understanding threshold for the dual-role cooperation rules upon first encounter, but it also prompts players to actively search for mechanics cues.

As shown in Figure 4.(b), the themes are concentrated and largely consistent with the CBT objectives. The "Perfectionism" level was primarily interpreted as "Don't persue perfection" (70%) and "Don't be tempted" (15%). "Overgeneralization" was mainly interpreted as "Perseverance means victory" (45%) and "I won't always fail" (35%). The "Jumping to Conclusions" level was primarily distilled as "Try first and then draw a conclusion" (55%). The "Magnification" level was mainly interpreted as "Don't magnify fear" (65%) and "Pay attention to illusions" (30%). The most consistent interpretation of the "Personalization" level was "Don't bear everything alone" (85%). Overall, this shows that players can reliably translate the mechanics into clear psychological meanings.

As shown in Figure 4.(c), realistic resonance mainly occurred in high-frequency student scenarios. "Perfectionism" was more often associated with academic pressure, such as college entrance exam pressure (70%) and university final exam pressure (15%). "Overgeneralization" was difficult for many participants to relate to in real-world experiences (35%), with other relevant experiences mainly being exam or competition failures. "Jumping to Conclusions" were more likely to be drawn from interpersonal interpretation (guessing a friend's thoughts, 35%) and self-assessment of abilities (learning a skill, 30%); "Magnification" was concentrated in social evaluation anxiety, such as making mistakes in school(30%), not receiving a timely response (30%), and saying the wrong thing (15%). The distribution of "Personalization" was relatively even, for example, team losed the match (20%), seeing others unhappy (15%), and parents arguing (15%).

Overall results indicate that the "World+You-World" level mechanism not only effectively triggers initial reactions similar to the distortion type (Figure 4.(a)), but is also consistently interpreted by players as corresponding CBT information (Figure 4.(b)), and generates transfer associations in real life such as academics and interpersonal relationships (Figure 4.(c)), providing qualitative evidence for the comprehensibility and transferability of TPR-MMF.

# 4. Discussion

This pilot study developed a serious mental health game, "World + You - World," based on the proposed TPR-MMF, aiming to address the low student participation and high level of boredom in traditional serious mental health education games. The results positively responded to the two core research questions, validating the effectiveness and feasibility of transforming cognitive behavioral therapy (CBT) mechanisms into an implicit game program.

Regarding RQ1, IMI data showed that compared to traditional explicit educational dialogue games, the game designed based on TPR-MMF significantly improved in dimensions such as interest/enjoyment, perceived competence, perceived choice, and value/usefulness, exhibiting a larger effect size. This indicates that abandoning didactic psychological knowledge indoctrination and deeply integrating CBT goals into the core loop of action-based games effectively aligns with students' entertainment preferences, thereby significantly enhancing the appeal of the intervention. Players gained a greater sense of autonomy and competence during the experience, which not only significantly improved the compliance problems often faced by traditional digital CBT but also provided a strong intrinsic motivation for sustained mental health support.

Qualitative analysis of RQ2 revealed that players initially exhibited intuitive strategies highly similar to cognitive distortion when faced with challenges related to "perfectionism" and "overgeneralization." The game, through its "expectation violation" mechanism, induced cognitive dissonance, successfully forcing players to abandon fixed thinking patterns and explore new paths. More importantly, the vast majority of players accurately interpreted the psychological metaphors behind the mechanism (such as "fear is often amplified" and "don't bear everything alone"), spontaneously resonating with high-frequency anxiety situations they faced in real life, such as academic pressure and interpersonal relationships. This embodied "behavioral validation" demonstrates the unique advantage of procedural rhetoric in mental health education: reflection evoked through interactive experiences is more profound than mere textual reading.

While this pilot study achieved positive preliminary results, certain limitations remain. First, the sample size was relatively small (N=28), and the participants were primarily university students, which somewhat limits the generalization of the findings to a wider student population. Second, the current evaluation mainly focuses on immediate motivation and experience after a single intervention, lacking long-term longitudinal tracking of participants' anxiety or stress symptoms. Future research should expand the sample size and incorporate long-term follow-up to further explore the long-term effects of such implicit mechanisms of serious games on mental health education.

## 5. Conclusions

This pilot study proposes a CBT-informed Serious Game Design Framework (TPR-MMF) centered on participatory experience, and provides an innovative solution for mental health education for students by developing and evaluating the game prototype "World+You-World". The study confirms that transforming the cognitive restructuring goals of CBT into implicit game mechanics and procedural rhetoric can significantly improve players' intrinsic motivation and experience quality. Simultaneously, this design effectively guides players to develop profound psychological awareness during game interaction and transfer this awareness to stressful situations in real life. This study significantly improves the "boring" nature of traditional mental health education materials, providing a highly attractive and low-barrier design paradigm for future digital mental health education products, and is expected to play a significant role in preventing student mental health problems.

## Data availability

The data supporting the results of this study can be provided by the corresponding author upon reasonable request. As these data include personal mental health information of participants, they will not be publicly released to protect their privacy.

## References

Benton, T. D., Boyd, R. C., & Njoroge, W. F. (2021). Addressing the global crisis of child and adolescent mental health. JAMA Pediatrics, 175(11), 1108–1110.

Boucher, E. M., & Raiker, J. S. (2024). Engagement and retention in digital mental health interventions: A narrative review. BMC Digital Health, 2(1), 52.

Carvajal-Velez, L., Requejo, J. H., Ahs, J. W., Idele, P., Adewuya, A., Cappa, C., ... & Kohrt, B. A. (2023). Increasing data and understanding of adolescent mental health worldwide: UNICEF's measurement of mental health among adolescents at the population level initiative. Journal of Adolescent Health, 72(1), S12-S14.

Christ, C., Schouten, M. J. E., Blankers, M., van Schaik, D. J. F., Beekman, A. T. F., Wisman, M. A., Stikkelbroek, Y. A. J., & Dekker, J. J. M. (2020). Internet and computer-based cognitive behavioral therapy for anxiety and depression in adolescents and young adults: Systematic review and meta-analysis. Journal of Medical Internet Research, 22(9), e17831.

David, O. A., Costescu, C., Cardos, R., & Mogoaşe, C. (2020, December). How effective are serious games for promoting mental health and health behavioral change in children and adolescents? A systematic review and meta-analysis. In child & youth care forum (Vol. 49, No. 6, pp. 817-838). New York: Springer US.

De Jaegere, E., van Heeringen, K., Emmery, P., Mommerency, G., & Portzky, G. (2024). Effects of a serious game for adolescent mental health on cognitive vulnerability: Pilot usability study. JMIR Serious Games, 12, e47513.

Dongjun, Z., Mingyue, W., Xinqi, L., Lina, W., Jiali, W., & Mengyao, J. (2025). Trends in depressive and anxiety disorders among adolescents and young adults (aged 10–24) from 1990 to 2021: a global burden of disease study analysis. Journal of Affective Disorders, 387, 119491.
Fleming, T. M., Bavin, L., Stasiak, K., Hermansson-Webb, E., Merry, S. N., Cheek, C., Lucassen, M., Lau, H. M., Pollmuller, B., & Hetrick, S. (2017). Serious games and gamification for mental health: Current status and promising directions. Frontiers in Psychiatry, 7, 215.
Fleming, T., Lucassen, M., Stasiak, K., Sutcliffe, K., & Merry, S. (2021). Technology Matters: SPARX–computerised cognitive behavioural therapy for adolescent depression in a game format. Child and Adolescent Mental Health, 26(1), 92-94.
García-Cabrera, E., Luna-Perejón, F., Pertegal-Vega, M. Á., Munoz-Saavedra, L., Sevillano-Ramos, J. L., & Miró-Amarante, L. (2025). Video game player profiles among university students: Impact of game preferences and academic background. Computers and Education Open, 100280.
Gómez-León, M. I. (2025). Serious games to support emotional regulation strategies in educational intervention programs with children and adolescents: Systematic review and meta-analysis. Heliyon, 11(4).
Hammady, R., & Arnab, S. (2022). Serious gaming for behaviour change: A systematic review. Information, 13(3), 142.
Hsu, T. C. (2026). A Generative AI Driven Interactive Narrative Serious Game for Stress Relief and Its Randomized Controlled Pilot Study. SSRN. https://papers.ssrn.com/sol3/papers.cfm?abstract_id=6471398
Jungbluth, N. J., & Shirk, S. R. (2013). Promoting homework adherence in cognitive-behavioral therapy for adolescent depression. Journal of Clinical Child & Adolescent Psychology, 42(4), 545–553.
Kazantzis, N. (2021). Introduction to the special issue on homework in cognitive behavioral therapy: New clinical psychological science. Cognitive Therapy and Research, 45(2), 205–208.
Li, H. (2019). A side scrolling game based on Unity3D. Practical Electronics, (16), 46–48, 92.
Linardon, J., Messer, M., Reid, R., Bolger, T., & Andersson, G. (2025). Absolute and relative rates of treatment non-initiation, dropout, and attrition in internet-based and face-to-face cognitive-behavioral therapy: a meta-analysis of randomized controlled trials. Cognitive Behaviour Therapy, 1-14.
Liu, Z., & Kuai, M. (2025). The global burden of depression in adolescents and young adults, 1990–2021: Systematic analysis of the global burden of disease study. BMC Psychiatry, 25(1), 767.
Lu, B., Lin, L., & Su, X. (2024). Global burden of depression or depressive symptoms in children and adolescents: A systematic review and meta-analysis. Journal of Affective Disorders, 354, 553–562.
Merry, S. N., Stasiak, K., Shepherd, M., Frampton, C., Fleming, T., & Lucassen, M. F. (2012). The effectiveness of SPARX, a computerised self help intervention for adolescents seeking help for depression: randomised controlled non-inferiority trial. Bmj, 344.
Michaud, P. A., Visser, A., Vervoort, J. P., Kocken, P., Reijneveld, S. A., & Jansen, D. E. (2020). Availability and accessibility of primary mental health services for adolescents: an overview of national recommendations and services in EU. European journal of public health, 30(6), 1127-1133.
Oud, M., de Winter, L., Vermeulen-Smit, E., Bodden, D., Nauta, M., Stone, L., van den Heuvel, M., Al Taher, R., de Graaf, I., Kendall, T., Engels, R., & Stikkelbroek, Y. (2019). Effectiveness of CBT for children and adolescents with depression: A systematic review and meta-regression analysis. European Psychiatry, 57, 33–45.
Pegg, S., Hill, K., Argiros, A., Olatunji, B. O., & Kujawa, A. (2022). Cognitive behavioral therapy for anxiety disorders in youth: efficacy, moderators, and new advances in predicting outcomes. Current Psychiatry Reports, 24(12), 853-859.
Schlack, R., Peerenboom, N., Neuperdt, L., Junker, S., & Beyer, A. K. (2021). The effects of mental health problems in childhood and adolescence in young adults: Results of the KiGGS cohort. Journal of health monitoring, 6(4), 3.
Schoneveld, E. A., Malmberg, M., Lichtwarck-Aschoff, A., Verheijen, G. P., Engels, R. C., & Granic, I. (2016). A neurofeedback video game (MindLight) to prevent anxiety in children: A randomized controlled trial. Computers in Human Behavior, 63, 321-333.
Schoneveld, E. A., Wols, A., Lichtwarck-Aschoff, A., Otten, R., & Granic, I. (2020). Mental health outcomes of an applied game for children with elevated anxiety symptoms: a randomized controlled non-inferiority trial. Journal of Child and Family Studies, 29(8), 2169-2185.
Shah, K. (2024). How American adults and teens differ in their gaming choices. YouGov. https://yougov.com/en-us/articles/48499-how-american-adults-and-teens-differ-in-their-gaming-choices
Silva, S. A., Silva, S. U., Ronca, D. B., Gonçalves, V. S. S., Dutra, E. S., & Carvalho, K. M. B. (2020). Common mental disorders prevalence in adolescents: A systematic review and meta-analyses. PloS one, 15(4), e0232007.
Townsend, C., Humpston, C., Rogers, J., Goodyear, V., Lavis, A., & Michail, M. (2022). The effectiveness of gaming interventions for depression and anxiety in young people: systematic review and meta-analysis.

BJPsych Open, 8(1), e25.
Uhlhaas, P. J., Davey, C. G., Mehta, U. M., Shah, J., Torous, J., Allen, N. B., ... & Wood, S. J. (2023). Towards a youth mental health paradigm: a perspective and roadmap. Molecular psychiatry, 28(8), 3171-3181.
Wahdi, A. E., Astrini, Y. P., Setyawan, A., Fine, S. L., Ramaiya, A., Li, M., Wado, Y. D., Loi, V. M., Maravilla, J. C., Scott, J. G., Wilopo, S. A., & Erskine, H. E. (2025). Mental health service use among adolescents in three low- and middle-income countries: An analysis of the National Adolescent Mental Health Surveys. Child and Adolescent Psychiatry and Mental Health, 19(Suppl. 1), 84.
Zeiler, M., Vögl, S., Prinz, U., Werner, N., Wagner, G., Karwautz, A., Zeller, N., Ackermann, L., & Waldherr, K. (2025). Game design, effectiveness, and implementation of serious games promoting aspects of mental health literacy among children and adolescents: Systematic review. JMIR Mental Health, 12(1), e67418.
Zhang, X., Liang, Z., & Kim, Y. (2025). Effectiveness of school-based cognitive behavioral therapy in alleviating anxiety and depression among high-risk child and young people: A Bayesian meta-analysis with meta-regression. Cognitive Therapy and Research, 1–12.

# Appendix A

## Appendix A.1 Intrinsic Motivation Inventory

Interest/Enjoyment
I enjoyed doing this activity very much
This activity was fun to do.
I thought this was a boring activity. (R)
This activity did not hold my attention at all. (R)
I would describe this activity as very interesting.
I thought this activity was quite enjoyable.
While I was doing this activity, I was thinking about how much I enjoyed it.
Perceived Competence
I think I am pretty good at this activity.
I think I did pretty well at this activity, compared to other students.
After working at this activity for a while, I felt pretty competent.
I am satisfied with my performance at this task.
I was pretty skilled at this activity.
This was an activity that I couldn't do very well. (R)
Effort/Importance
I put a lot of effort into this.
I didn't try very hard to do well at this activity. (R)
I tried very hard on this activity.
It was important to me to do well at this task.
I didn't put much energy into this. (R)
Pressure/Tension
I did not feel nervous at all while doing this. (R)
I felt very tense while doing this activity.
I was very relaxed in doing these. (R)
I was anxious while working on this task.
I felt pressured while doing these.
Perceived Choice
I believe I had some choice about doing this activity.
I felt like it was not my own choice to do this task. (R)
I didn't really have a choice about doing this task. (R)
I felt like I had to do this. (R)
I did this activity because I had no choice. (R)
I did this activity because I wanted to.
I did this activity because I had to. (R)
Value/Usefulness

I believe this activity could be of some value to me.
I think that doing this activity is useful for building positive cognition.
I think this is important to do because it can building positive cognition.
I would be willing to do this again because it has some value to me.
I think doing this activity could help me to building positive cognition.
I believe doing this activity could be beneficial to me.
I think this is an important activity.